\documentclass[journal=jacsat,manuscript=article,layout=onecolumn]{achemso}

\setkeys{acs}{keywords = true}
\setkeys{acs}{maxauthors = 0} 
\setkeys{acs}{articletitle = true}

\usepackage{chemformula} 
\usepackage[T1]{fontenc} 

\author{Zhao Liu}
\email{zhaoliu@swin.edu.au}
\affiliation{Centre of Quantum Technology Theory, Swinburne University of Technology, Melbourne, Victoria 3122, Australia}
\alsoaffiliation{Department of Materials Science and Engineering, Monash University, Victoria 3800, Australia}
\alsoaffiliation{ARC Centre of Excellent in Future Low-Energy Electronics Technologies, Monash University, Victoria 3800, Australia}

\author{Nikhil V. Medhekar}
\email{Nikhil.Medhekar@monash.edu}
\affiliation{Department of Materials Science and Engineering, Monash University, Victoria 3800, Australia}
\alsoaffiliation{ARC Centre of Excellent in Future Low-Energy Electronics Technologies, Monash University, Victoria 3800, Australia}

\title[An \textsf{achemso} demo]
  {$d$-Wave Polarization-Spin Locking in Two-Dimensional Altermagnets}


\begin{document}

\begin{tocentry}

\includegraphics{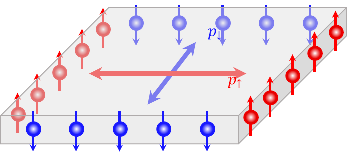}

\end{tocentry}


\begin{abstract}
We report the emergence of an uncharted phenomenon, termed $d$-wave polarization-spin locking (PSL), in two-dimensional (2D) altermagnets. This phenomenon arises from nontrivial Berry connections, resulting in perpendicular electronic polarizations in the spin-up and spin-down channels. Symmetry-protected $d$-wave PSL occurs exclusively in $d$-wave altermagnets with tetragonal layer groups. To identify 2D altermagnets capable of exhibiting this phenomenon, we propose a symmetry-eigenvalue-based criterion, and a rapid method by observing the spin-momentum locking. Using first-principles calculations, monolayer Cr$_2$X$_2$O (X = Se, Te) characterizes promising candidates for $d$-wave PSL, driven by the unusual charge order in these monolayers. This unique polarization-spin interplay leads to spin-up and spin-down electrons accumulating at orthogonal edges, enabling potential applications as spin filters or splitters in spintronics. Furthermore, $d$-wave PSL introduces an unexpected spin-driven ferroelectricity in conventional antiferromagnets. Such magnetoelectric coupling positions $d$-wave PSL as an ideal platform for fast antiferromagnetic memory devices. Our findings not only expand the landscape of altermagnets, complementing conventional collinear ferromagnets and antiferromagnets, but also highlight tantalizing functionalities in altermagnetic materials, potentially revolutionizing information technology.  
\end{abstract}

\maketitle
The discovery of lifted Kramers degeneracy in collinear antiferromagnets (AFs) without the aid of spin-orbit coupling (SOC) \cite{Noda2016, Hayami2019, Hayami2020, Smejkal2020, Yuan2020, Egorov2021, Mazin2021, Ma2021} has fueled the rapid development of altermagnetsim (AM) \cite{Smejkal2022}. Initially regarded as unconventional symmetry broken phases due to Fermi liquid instabilities in the spin channel \cite{Wu2007}, now AMs has been extended to insulating systems as well under the framework of nontrivial spin space group \cite{Brinkman1966, Litvin1974, Xiao2024, Chen2024, JiangY2024, ZengS2024}, whose group element can be expressed as $[\mathcal{R}_s||\mathcal{R}_l]$ with transformation $\mathcal{R}_s$ and $\mathcal{R}_l$ acting solely in the spin and real space. Recently, 23 two-dimensional (2D) and 221 three-dimensional (3D) AM candidates have been tabulated \cite{Bai2024}, among which lifted Kramers degeneracy has been observed in MnTe \cite{Krempasky2024, Lee2024, Souma2024}, CrSb \cite{Reimers2024, Zeng2024, Ding2024, Yang2025}, and RuO$_2$ under scrutiny \cite{Feng2022, Bose2022, Bai2022, Karube2022, Wang2024, Liao2024, Fedchenko2024, LinZ2024, Hiraishi2024, LiuJ2024, Wenzel2025, Qian2025}. In addition to spin splitting, AMs exhibit various exotic physical phenomena related to time-reversal symmetry breaking, including spin transport properties \cite{Shao2021, Smejkal2022-2, Jin2024, Das2024}, (noncollinear) spin current generation \cite{Naka2019, Naka2021, Ma2021,  Hernandez2021, Wu2024}, chiral magnons \cite{Smejkal2023, Cui2023, Sodequist2024, Liu2024}, magnetic multipoles \cite{Suzuki2019, Bhowal2024}, interplay with superconductivity \cite{Mazin2022, Zhu2023, Brekke2023, Chakraborty2024, ZhangS2024}, etc. These remarkable discoveries establish a firm foundation for AMs with versatile functionalities in spintronics, magnetism, and superconductivity \cite{Bai2024, Smejkal2022-3, Song2025}. 

\begin{figure}
	\includegraphics[width=0.50\linewidth]{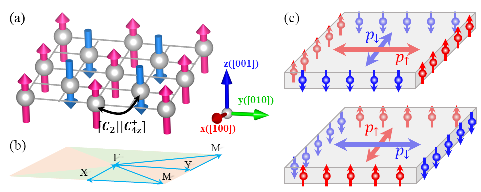}
	\caption{(a) A schematic of 2D $d$-wave altermagnet in the $xy$ plane. The red and blue arrows represent the two spin sublattices connected by transformation $[C_2||C^+_{4z}]$. (b) First Brillouin zone with four TRIM $\Gamma = (0, 0)$, X = $(\frac{1}{2},0)$, Y=$(0, \frac{1}{2})$ and M = $(\frac{1}{2},\frac{1}{2})$. The green/orange colored regions represent opposite spin splitting. (c) Two configurations of $d$-wave PSL. Top/bottom: spin-up and down electronic polarizations are along $y$/$x$ and $x$/$y$ direction, with vanishing ionic polarization, spin-up and down electrons (red and blue balls) accumulate at surfaces perpendicular to $y$/$x$ and $x$/$y$ direction, respectively. 
	} \label{fig:sym}
\end{figure}

AMs can also contribute uniquely to Berry phase physics \cite{Berry1984}. In crystals with time-reversal symmetry breaking, the Bloch states acquire nonvanishing Berry curvatures, the geometric analog of the magnetic field in momentum space \cite{Xiao2010}. The spin-momentum locking nodal surface in AMs allows nonzero Berry curvatures under SOC and gives birth to the anomalous Hall effect \cite{Smejkal2020, Mazin2021, Naka2020, Naka2022, Smejkal2022-4, Betancourt2023, Reichlova2024, Leiviska2024, Han2024, Zhou2025, Sato2024, Tan2024, Fang2024, Zhou2024}, an effect absent in conventional collinear AFs. However, nonzero Berry curvature is unnecessary for the nontrivial topology, as demonstrated by the Aharonov-Bohm effect \cite{Aharonov1959}. Berry connection, the geometric analog of the vector potential in momentum space, can also lead to nontrivial topology. According to the modern theory of polarization \cite{Resta1994}, the integration of Berry connection over the first Brillouin zone yields electronic polarization $\vec{p}_{ele}$, which can serve as a topological invariant even with zero Berry curvature everywhere \cite{Liu2017, Jadaun2013, LiY2024}. Although Berry curvature-induced anomalous Hall effect has been disclosed in AMs, the intrinsic integration between Berry connection-induced quantized electronic polarization and AF order has not yet been reported. 

Here we construct a minimal model based on layer group $G = P\frac{4}{m}$. By incorporating the magnetic order demonstrated in {\color{blue} Fig.\ref{fig:sym}(a)}, a nontrivial spin layer group describing AM can be constructed:
\begin{equation}
\mathcal{R}= [E||H] + [C_2||C^+_{4z} H]
\end{equation}
where $C_2$ is the spin flip symmetry, $C^+_{4z}$ is the four-fold counterclockwise rotation along $z$ direction. $H = Pmmm = \{E, \mathcal{P}, C_{2z},  M_{z} \}$ ($E/C/M$ is the identity/rotation/mirror transformation) is the halving subgroup of $G$ according to the coset partition $G = H \bigoplus C^+_{4z} H$. $\mathcal{R}$ is labeled as $P\frac{4'}{m}$ in magnetic space group notation and $\frac{^2{4}}{^1{m}}$ in Litvin's notation \cite{Litvin1974}. As shown in {\color{blue} Fig.\ref{fig:sym}(a)}, the transformations in $[E||H]$ only interchange atoms within one spin sublattice, while $[C_2||C^+_{4z} H]$ interchange both spin sublattices. 

The spin layer group $\mathcal{R}$ endows a unique spin texture called $d$-wave spin-momentum locking (SML). We denote the Bloch state and its eigenvalue as $|\phi(\vec{k}, \sigma) \rangle$ and $E(\vec{k}, \sigma)$, where $\vec{k}$ is the momentum ($k_z = 0$ and is ignored herein) and $\sigma = \pm 1$ represent up/down spins. First, for $\vec{k}$ whose little group does not contain element in $AH$, such as the $\Gamma$-X and $\Gamma$-Y paths, spin degeneracy is lifted, i.e. $E(\vec{k}, \sigma) \neq E(\vec{k}, -\sigma)$. Second, $[C_2||C^+_{4z}]$ transforms $E(\vec{k}, \sigma)$ as (disregarding translation symmetry for simplicity):
\begin{equation}
[C_2||C^+_{4z}] E(k_x, k_y, \sigma) = E(k_y, -k_x, -\sigma)
\end{equation}
For $\vec{k}$ whose little group contains $[C_2||C^+_{4z}]$, spin degeneracy is guaranteed like $\Gamma$ and $M$ points. Meanwhile, the little groups of $\Gamma$-X and $\Gamma$-Y paths do not contain $[C_2||C^+_{4z}]$ and we have $E(\Gamma \rightarrow X, \sigma) = E(\Gamma \rightarrow Y, -\sigma)$. Together, these symmetries result in $d$-wave SML, as demonstrated at the bottom of {\color{blue} Fig.\ref{fig:sym}(b)}. 

The existence of $\mathcal{P}$ allows Bloch states at time-reversal invariant momentum (TRIM) $\kappa$ to have either even or odd parity: $\eta(|\phi(\kappa, \sigma) \rangle)=\pm 1$, and denoting $\kappa_{-, \sigma}$ as the number of occupied Bloch states with odd parity in spin channel $\sigma$, the electric polarization $(p_{ele,x,\sigma}, p_{ele,y,\sigma})$ in insulators can be calculated as follows\cite{Fang2012, Turner2012, Po2017, Miert2018, LiuZ2024}:
\begin{equation}
(p_{ele,x,\sigma}, p_{ele,y,\sigma}) = (\frac{\Gamma_{-,\sigma} - X_{-,\sigma}}{2}, \frac{\Gamma_{-,\sigma} - Y_{-,\sigma}}{2}) \mod{1}   
\end{equation}
Since $[[E||P],[C_2||C^+_{4z}]]_-$ = 0, we have $\eta(|\phi(\Gamma, \sigma)\rangle)=\eta(|\phi(\Gamma, -\sigma)\rangle)$ and $\eta(|\phi(X, \sigma)\rangle)=\eta(|\phi(Y, -\sigma)\rangle)$, so the polarization satisfies:
\begin{equation}
p_{ele,x,\sigma} = p_{ele,y,-\sigma}
\label{eq:PSL-1}
\end{equation}
At the same time, the absence of $C^+_{4z}$ in $H$ generally gives $\eta(|\phi(X, \sigma)\rangle) \neq \eta(|\phi(Y, \sigma)\rangle)$ and thus $X_{-, \sigma} \neq Y_{-,\sigma}$. If $X_{-, \sigma}$ and $Y_{-, \sigma}$ have different parity, i.e. 
\begin{equation}
X_{-, \sigma} +  Y_{-, \sigma} = 1 \mod{2}  
\label{eq:PSL-2}
\end{equation}
then $p_{ele,x,\sigma} \neq p_{ele,y,\sigma}$ is ensured regardless of $\Gamma_-$. The total polarization $\vec{p}$, the summation of ionic polarization $\vec{p}_{ion}$ and $\vec{p}_{ele}$, is gauge invariant and thus experimental observable. With vansihing $\vec{p}_{ion}$, $\vec{p}$ is totally contributed by $\vec{p}_{ele}$ and {\color{blue} Eq.\ref{eq:PSL-1}} and {\color{blue} Eq.\ref{eq:PSL-2}} combined give perpendicular $\vec{p}_{\sigma}$ and $\vec{p}_{-\sigma}$, as schematically plotted in {\color{blue} Fig.\ref{fig:sym}(c)}. Such an anisotropic polarization in both real and spin space resembles that of $d$-wave SML, therefore it is named $d$-wave polarization-spin locking here. In fact, {\color{blue} Eq.\ref{eq:PSL-2}} is the necessary and sufficient condition for $d$-wave PSL because $\vec{p}_{\sigma} $ and $\vec{p}_{-\sigma}$ will be identical if $X_{-, \sigma}$ and $Y_{-,\sigma}$ have the same parity.

The above analysis enables the identification of 2D AM candidates hosting $d$-wave PSL. Firstly, we are interested in the $d$-wave PSL protected by symmetry, excluding accidental $d$-wave PSL in layer groups such as $P1$. Among the seven crystal systems, only the tetragonal system satisfies {\color{blue} Eq.\ref{eq:PSL-1}}. For tetragonal layer groups without $\mathcal{P}$, a criterion analogous to {\color{blue} Eq.\ref{eq:PSL-2}} can be derived based on symmetry $C_{2z}$  (see Section A of Supplementary Information ). Secondly, while tetragonal systems can host both $d$-wave and $g$-wave SML, it can be further demonstrated that only $d$-wave SML shown in {\color{blue} Fig.\ref{fig:sym}(b)} can exhibit $d$-wave PSL (see Section B of Supplementary Information). This provides a rapid approach for screening $d$-wave PSL, as SML is easily obtained. Finally, the above analysis can be naturally extended from 2D to 3D systems, where planar $d$-wave PSL is possible \cite{Turner2012}. 

\begin{figure*}
	\includegraphics[width=1.00\linewidth]{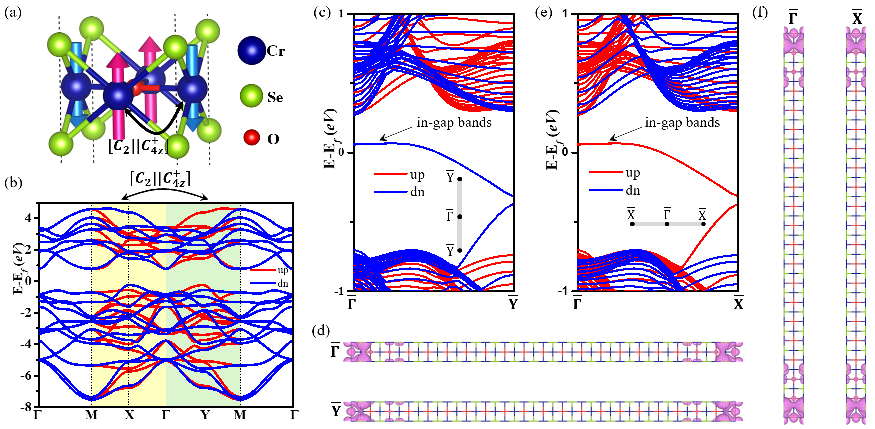}
	\caption{(a) Atomic configuration of monolayer Cr$_2$Se$_2$O. The two spin sublattices are connected by $[C_2||C^+_{4z}]$. (b) Spin-resolved band structure along high symmetric $\vec{k}$-path. Under $[C_2||C^+_{4z}]$, the spin-up (down) bands along M-X-$\Gamma$ (yellow colored region) are transformed to spin-down (up) bands along M-Y-$\Gamma$ (green colored region). Spin-resolved band structure for a ribbon with length $L_x$ = 20.5 a$_0$ along (c) $x$ and (e) $y$ direction, insert is the 1D FBZ. The charge density distribution for the in-gap Bloch states at (d) $\bar{\Gamma}$ and $\bar{Y}$ and (e) $\bar{\Gamma}$ and $\bar{X}$ point.
	} \label{fig:CrSeO}
\end{figure*}

2D AM candidates, including V$_2$Se$_2$O/V$_2$Te$_2$O,  Cr$_2$Se$_2$O/Cr$_2$Te$_2$O and Fe$_2$Se$_2$O, are predicted to host spin layer group with $\mathcal{R}$ as its subgroup \cite{Bai2024}. By doping Rb into bulk V$_2$Te$_2$O and K into bulk V$_2$Se$_2$O, the 2D AM nature has been reported recently \cite{Zhang2024, Jiang2024}. Via applying {\color{blue} Eq.\ref{eq:PSL-2}} to the 2D AMs exhausted by Bai et al. \cite{Bai2024}, monolayer Cr$_2$X$_2$O (X = Se, Te) is screened out as a promising candidate for $d$-wave PSL (see Section D of Supplementary Information). To illustrate, we focus on monolayer Cr$_2$Se$_2$O. It is noteworthy that different atomic configurations correspond to different gauge fixing conditions, we choose the atomic configuration in {\color{blue} Fig.\ref{fig:CrSeO}(a)}, so $\vec{p}_{ion} = 0$ and $\vec{p} = \vec{p}_{ele}$. The spin-resolved band structure are shown in {\color{blue} Fig.\ref{fig:CrSeO}(b)}, and as expected, the bands along the two paths (yellow- and green-colored region in {\color{blue} Fig.\ref{fig:CrSeO}(b)}) connected by $[C_2||C^+_{4z}]$ exhibit $d$-wave SML, in accordance with previous results \cite{Gong2024}. The insulating nature in {\color{blue} Fig.\ref{fig:CrSeO}(b)} allows for a well-defined electronic polarization. {\color{blue} Tab.\ref{tab:parity}} lists the parity distribution for all the 30 valance bands of both spins at the four TRIM. Clearly, AM monolayer Cr$_2$Se$_2$O satisfies the criterion {\color{blue} Eq.\ref{eq:PSL-2}}, and we have $(p_{ele, x,\uparrow}, p_{ele, y,\uparrow}) = (0, \frac{1}{2})$, $(p_{ele, x,\downarrow}, p_{ele, y,\downarrow}) = (\frac{1}{2}, 0)$, which corresponds to the $d$-wave PSL configuration at the top of {\color{blue} Fig.\ref{fig:sym}(b)}. Such a result is also confirmed by first-principle calculations based on the Berry phase. 

\begin{table}[b]
	\caption{Parity eigenvalues of all occupied bands at the four TRIM of monolayer Cr$_2$Se$_2$O in different spin channels.}
	\label{tab:parity}
	\begin{tabular}{c|c|c|c|c}
		\hline\hline
		TRIM  & $(+,\uparrow)$ & $(-,\uparrow)$ & $(+,\downarrow)$ &  $(-,\downarrow)$ \\
		\hline
		$\Gamma$ & 8  & 7 & 8 & 7 \\
		$X$      &10  & 5 & 7 & 8 \\
		$Y$      & 7  & 8 &10 & 5 \\
		$M$      & 5  &10 & 5 &10 \\
		\hline\hline
	\end{tabular}
\end{table}

When there is nonvanishing total polarization, one physical observable is the accumulation of charges at the open boundary, resulting in edge states inside the bulk gap. In contrast with traditional edge states where spin-up and down electrons both assemble at the same edge, here $d$-wave PSL can break such a spin degeneracy. To illustrate this, $\mathcal{P}$-symmetric ribbons with a length of 20.5 a$_0$ (a$_0$ is the in-plane lattice constant) are constructed along the $x$ and $y$ directions. {\color{blue} Fig.\ref{fig:CrSeO}(c)} and {\color{blue} Fig.\ref{fig:CrSeO}(e)} show the spin-resolved band structures. For a ribbon along $x$ ($y$) direction, only spin-down (up) bands exist within the bulk gap. Each in-gap band is doubly degenerate due to $\mathcal{P}$, which can potentially yield non-Abelian statics. {\color{blue} Fig.\ref{fig:CrSeO}(d)} and {\color{blue} Fig.\ref{fig:CrSeO}(f)} show the spatial charge density distribution of one in-gap band at $\overline{\Gamma}$ and boundary momentum $\overline{X}$/$\overline{Y}$. With charges mainly localizing at the two edges, these in-gap bands are edge states as a result of nontrivial polarizations.

The reason why monolayer Cr$_2$Se$_2$O satisfy {\color{blue} Eq.\ref{eq:PSL-2}} stems from the unusual valence state Cr$^{2+}_2$Se$^{1-}_2$O$^{2-}$, rather than the formal valence state Cr$^{3+}_2$Se$^{2-}_2$O$^{2-}$. To see this, we divide the whole system into two independent spinless models (these two spinless components are still named spin-up and spin-down) and ignore the inner O $s$ and Se $s$ shells for simplicity. As shown in Section F of Supplementary Information, these two distinct valence states give the same parity distribution and thus identical electronic polarization $\vec{p}_{ele, \uparrow} = (\frac{1}{2}, 0), \vec{p}_{ele,\downarrow} = (0, \frac{1}{2})$. However, the $\vec{p}_{ion}$ differs, and only the valence state Cr$^{2+}_2$Se$^{1-}_2$O$^{2-}$ can give $d$-wave PSL. 

\begin{figure*}
	\includegraphics[width=1.00\textwidth]{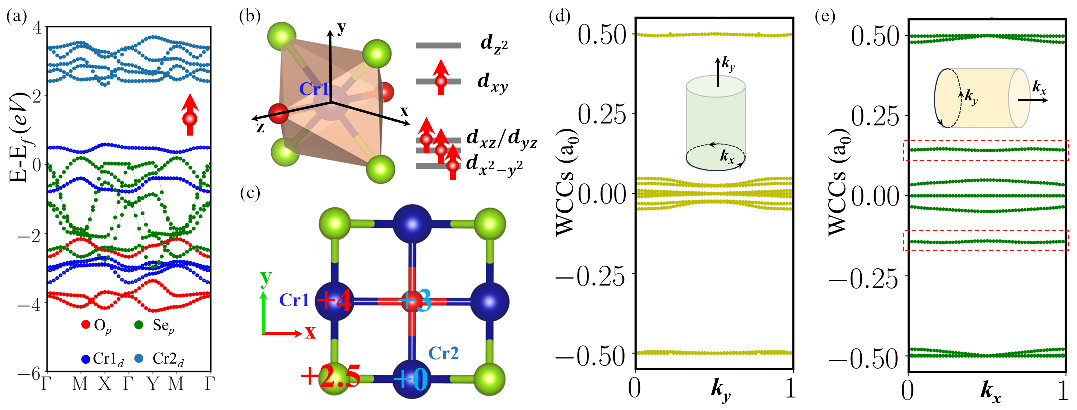}
	\caption{(a) Orbital-resolved band structure of Hamiltonian $\hat{H}_{Cr_d}+\hat{H}_{Se_p}+\hat{H}_{O_p}$ in the spin-up channel. (b) Ligand field splitting for the CrSe$_4$O$_2$ distorted octahedra. The $x$, $y$ and $z$ axes shown are local coordinate, to be distinguished from the global coordinate shown in (c). (c) Actual ionic charge in the spin-up channel (here the core ionic charge and electrons are ignored). (d)-(e) The evolution of the Wannier charge centers along $k_x$ and $k_y$ directions as indicated by the insert. 
	} \label{fig:charge}
\end{figure*}

To verify that Cr$^{2+}_2$Se$^{1-}_2$O$^{2-}$ is the actual valence state, we provide two evidences. Firstly we examine the intrinsic energies of Cr $d$, Se $p$, and O $p$ orbitals, which can be obtained by diagonalizing the block Hamiltonian $\hat{H}_{Cr_d}+\hat{H}_{Se_p}+\hat{H}_{O_p}$ ((see Section C of Supplementary Information)). {\color{blue} Fig.\ref{fig:charge}(a)} show the band structure in the spin-up channel. Obviously, the O $p$ shell has a much lower energy than both Se $p$ and Cr $d$ shells; therefore, the O $p$ shell is closed with a valence state $1-$ and an ionic charge +3 as marked in {\color{blue} Fig.\ref{fig:charge}(c)}. Since Cr2 has negative magnetization, its five $d$ orbitals are pushed to higher energy and contribute to the highest five conduction bands in {\color{blue} Fig.\ref{fig:charge}(a)}, correspondingly its ionic charge is 0 as shown in {\color{blue} Fig.\ref{fig:charge}(c)}. For Cr1, the splitting of the five $d$ bands directly originates from the crystal field splitting plotted in {\color{blue} Fig.\ref{fig:charge}(b)}. Although the highest $d$ band is empty, the second highest $d$ band is not, as there are Se $p$ bands above it. With seven empty bands in {\color{blue} Fig.\ref{fig:charge}(a)}, one band formed by Se $p$ orbitals must be empty, and the second highest $d$ is thus occupied as shown in {\color{blue} Fig.\ref{fig:charge}(b)}. Therefore, the actual valence state of Se is $1.5-$ as shown in {\color{blue} Fig.\ref{fig:charge}(c)}, to be distinguished from the formal valence state $2-$. Secondly, we consider 1D Wilson loop operators along $k_x$ and $k_y$ directions, as shown in the insert of {\color{blue} Fig.\ref{fig:charge}(d)-(e)}. The diagonalization of these operators gives the $x$- and $y$-directed Wannier charge centers (WCCs), indicating the electronic locations along the $x$ and $y$ directions. As plotted in {\color{blue} Fig.\ref{fig:charge}(d)-(e)}, there are 12 WCCs for the 12 valence bands and each WCC is quite flat with respect to $k_y$ and $k_x$. The summation of all these 12 WCCs is 0.5 a$_0$ and 0 along the $x$ and $y$ directions, corresponding to $\vec{p}_{ele,\uparrow} = (\frac{1}{2}, 0)$. In {\color{blue} Fig.\ref{fig:charge}(d)}, there are 9 WCCs around $x = 0$ a$_0$ and 3 WCCs around $x = 0.5$ a$_0$, being aligned with Cr1 + Se and O positions. Nevertheless, in {\color{blue} Fig.\ref{fig:charge}(e)}, despite the 6 WCCs around $y = 0.5$ a$_0$ and 4 WCCs around $y = 0$ a$_0$, there are 2 WCCs at around $y = \pm 0.14$ a$_0$ (red dashed rectangles) which are away from any atomic positions. Such a mismatch between WCCs and atomic positions indicates the strong hybridization between Cr1 $d_{xy}$ and Se $p_{x/y}$ orbitals in {\color{blue} Fig.\ref{fig:charge}(b)}, so the charges transfer from Se to Cr1 atoms. If no charge transfer occurs, the Se $p$ shell would be closed and contribute to 6 WCCs around $y = 0$ a$_0$, rather than 3 WCCs in {\color{blue} Fig.\ref{fig:charge}(e)}. 

Therefore, {\color{blue} Fig.\ref{fig:charge}(c)} lists the actual ionic charge which leads to $\vec{p}_{ion, \uparrow} = (\frac{1}{2}, \frac{1}{2}), \vec{p}_{ion, \downarrow} = (\frac{1}{2}, \frac{1}{2})$. Together with $\vec{p}_{ele, \uparrow} = (\frac{1}{2}, 0), \vec{p}_{ele, \downarrow} = (0, \frac{1}{2})$, we obtain the correct total polarization: $\vec{p}_{\uparrow} = (0, \frac{1}{2}), \vec{p}_{\downarrow} = (\frac{1}{2}, 0)$. Conversely, the ionic charge corresponds to the valence state Cr$^{3+}_2$Se$^{2-}_2$O$^{2-}$ yields $\vec{p}_{ion, \uparrow} = (\frac{1}{2}, 0), \vec{p}_{ion, \downarrow} = (0, \frac{1}{2})$, resulting in vanishing total polarizations.

\begin{figure}
	\includegraphics[width=0.50\linewidth]{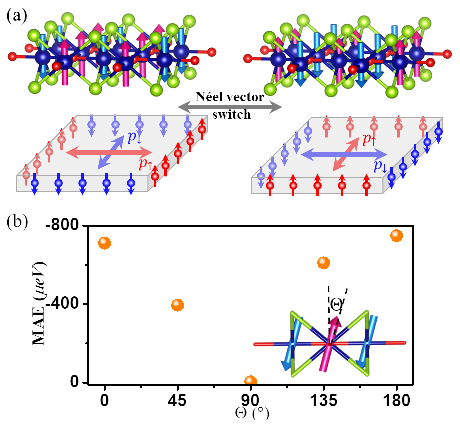}
	\caption{(a) Schematic 180$^{\circ}$ switching of N\'{e}el vector. (b) Energy evolution per unit cell with respect to magnetic moment rotation angle $\Theta$ in the $yz$ plane, as shown by the insert.
	} \label{fig:MAE}
\end{figure}

Finally, we discuss two potential applications of $d$-wave PSL. First, the spin-polarized edge states bestow monolayer Cr$_2$Se$_2$O with the ability to cooperate with spintronics and function as a spin filter or spin splitter. Secondly, spin-driven ferroelectricity provides an efficient way toward magnetoelectric coupling \cite{Tokura2014}, especially in AFs which can realize terahertz switching frequency with no stray fields and low damping spin current \cite{Baltz2018, Baierl2016}. Since $\mathcal{P}$ should be broken in both collinear AFs in the exchange striction model \cite{Picozzi2007, Tokura2014} and noncollinear AFs to induce spin chirality \cite{Katsura2005, Xiang2011, Zhang2017}, $d$-wave PSL represents a new type of spin-driven ferroelectricity without breaking $\mathcal{P}$. The 180$^{\circ}$ switch of N\'{e}el order in monolayer Cr$_2$Se$_2$O can induce a 90$^{\circ}$ planar rotation of polarization, as schematically shown in {\color{blue} Fig.\ref{fig:MAE}(a)}. These two configurations can work as binary '0' and '1', the basic logic bit for memory. Such a bit flip can be realized through either staggered N\'{e}el order spin-orbit torques \cite{Zelezny2014, Wadley2016} or non-staggered anti-dampling-like torques \cite{Moriyama2018, Chen2018}. To measure the ease of '0'-'1' switch in monolayer Cr$_2$Se$_2$O, the magnetic anisotropy energy is calculated. Using spin-orbit coupling as a perturbation, Whangbo et al. \cite{Whangbo2015} describe a method to predict the magnetic easy axis. Since the change in the magnetic quantum number between the highest occupied orbital $d_{xy}$ and lowest unoccupied orbital $d_{z^2}$ is 1 (see {\color{blue} Fig.\ref{fig:charge}(b)}), the magnetic moments prefer to order perpendicular to the Cr1-O bonds. {\color{blue} Fig.\ref{fig:MAE}(b)} confirms that the out-of-plane is the easy axis and the calculated energy barrier is around 0.8 meV/cell. This value is close to that in the AM candidate Mn$_5$Si$_3$, whose N\'{e}el order has been 180$^{\circ}$ switched via staggered N\'{e}el order spin-orbit torques \cite{Leiviska2024, Han2024} recently.

Two remarks are made before we conclude. First, the existence of nonvanishing polarization is just a necessary and insufficient conditions for ferroelectricity. Due to the existence of $\mathcal{P}$, the polarization in monolayer Cr$_2$Se$_2$O can not be switched in each spin channel as $(0, +\frac{1}{2})$ and $(0, -\frac{1}{2})$ are equivalent in the spin-up channel. In this sense, monolayer Cr$_2$Se$_2$O is not ferroelectrics. Nevertheless, it is plausible to switch the the polarization in the spin-up channel from $(0, \frac{1}{2})$ to $(\frac{1}{2}, 0)$ and simultaneously  from $(\frac{1}{2}, 0)$  to $(0, \frac{1}{2})$ in the spin-down channel, as elucidated in {\color{blue} Fig.\ref{fig:MAE}(a)}. This constitutes a new type of ferroelectricity driven by spin order. Second, we would like to compare our work with the previous work by Gong \textit{et al.} \cite{Gong2024} where monolayer Cr$_2$Se$_2$O is revealed as a 2D antiferromagnetic real Chern insulator. As a prior condition to be a real Chern insulator, the $\vec{p}_{ele}$ should vanish in both $x$ and $y$ directions \cite{Lee2020}. According to our results, monolayer Cr$_2$Se$_2$O has nonvansihing $\vec{p}_{ele}$, therefore, it is more appropriate to classify monolayer Cr$_2$Se$_2$O as nontrivial topological insulator characterized by quantized polarizations, which are coupled with spin.

In conclusion, we present a novel phenomenon termed $d$-wave PSL in two-dimensional altermagnets, and establish a criterion to identify potential material candidates. Monolayer Cr$_2$X$_2$O (X = Se, Te) are revealed to possess $d$-wave PSL, primarily driven by the unusual charge order. With promising applications in antiferromagnetic spintronics and memory devices, $d$-wave PSL is expected to draw immediate experimental attention.

\begin{suppinfo}

The Supporting Information is available free of charge on the ACS Publications website at: *** 

\begin{itemize}
  \item Criteria for $d$-wave PSL in different tetragonal layer groups; Fast screening of $d$-wave PSL by SML; Screening of known 2D altermagnetic insulator and details of first-principles calculations (PDF).
\end{itemize}

\end{suppinfo}

\begin{acknowledgement}
Z. L. thanks X. Yuan, and W. Zhu for helpful discussions. Z. L. and N. V. M. gratefully acknowledge the support from National Computing Infrastructure, Pawsey Supercomputing Facility and the Australian Research Council's Centre of Excellence in Future Low-Energy Electronic Technologies (CE170100039).
\end{acknowledgement}

\bibliography{MS-demo}

\end{document}